\RequirePackage{fix-cm}

\documentclass[twocolumn]{svjour3}          % twocolumn
\usepackage{lipsum}
\smartqed   %flush right qed marks, e.g. at end of proof
\usepackage{graphicx}
\usepackage{amsmath}
\usepackage{amsfonts}
\begin{document} \sloppy
%\lipsum[1-14]
\title{Coarse grid projection methodology: A partial mesh refinement tool for incompressible flow simulations
}



\author{A. Kashefi     %    \and
       % A. E. Staples %etc.
}


\institute{A. Kashefi \at
              Department of Mechanical Engineering, Stanford University, Stanford, CA 94305, USA \\
  %            Tel.: +123-45-678910\\
    %          Fax: +123-45-678910\\
              \email{kashefi@stanford.edu}             \\
  %         \and
           %A. E. Staples \at
           %   Faculty of Engineering, Engineering Science and Mechanics Program, Virginia Tech, Blacksburg, VA 24061, USA \\
  %            Tel.: +123-45-678910\\
    %          Fax: +123-45-678910\\
          %    \email{aestaples@vt.edu}           %  \\
}

\date{Short Communication}

\maketitle

\begin{abstract}
We discuss Coarse Grid Projection (CGP) methodology as a guide for partial mesh refinement of incompressible flow computations for the first time. Based on it, if for a given spatial resolution the numerical simulation diverges or the velocity outputs are not accurate enough, instead of refining both the advection-diffusion and the Poisson grids, the CGP mesh refinement suggests to only refine the advection-diffusion grid and keep the Poisson grid resolution unchanged. The application of the novel mesh refinement tool is shown in the cases of flow over a backward-facing step and flow past a cylinder. For the backward facing step flow, a three-level partial mesh refinement makes a previously diverging computation numerically stable. For the flow past a cylinder, the error of the viscous lift force is reduced from 31.501\% to 7.191\% (with reference to the standard mesh refinement results) by the one-level partial mesh refinement technique.
\end{abstract}

\section{Problem formulation}
\label{intro}

In order to simulate incompressible flows using pressure correction schemes [1], the computational cost on a given coarse grid with $N$ elements, $C_c$, is approximated by
\begin{equation}
C_c = C_v+C_p,
\end{equation}
where $C_v$ and $C_p$ comprise the numerical cost of the nonlinear advection-diffusion equation and the linear pressure Poisson equation, respectively. Now, if the two-dimensional coarse grid is uniformly refined by $l$-level, the simulation using a high-resolution grid with $M$ elements takes time $C_f$, roughly determined as
\begin{equation}
C_f \approx 4^l C_v+4^lC_p,
\end{equation}
where $4^l$ is a factor for cost scaling of the advection-diffusion and the Poisson equations in a two-dimensional problem. According to the CGP technique [2--8], the advection-diffusion equation is executed on the fine grid with $M$ elements, while the pressure Poisson equation is still solved on the coarse grid with $N$ elements. Generally, we show the resolution of a CGP simulation in the form of $M:N$, where $M$ and $N$ are defined as above. Hence, the computational cost of CGP, $C_{cgp}$, is estimated by
\begin{equation}
C_{cgp} \approx 4^lC_v+C_p+C_m,
\end{equation}
where $C_m$ is the mapping cost and is negligible in comparison with the other two terms of Eq. (3). One might see Sect. 2.3 of Ref. [7] for further details.

\begin{figure*}
\centering
\includegraphics[width=182 mm]{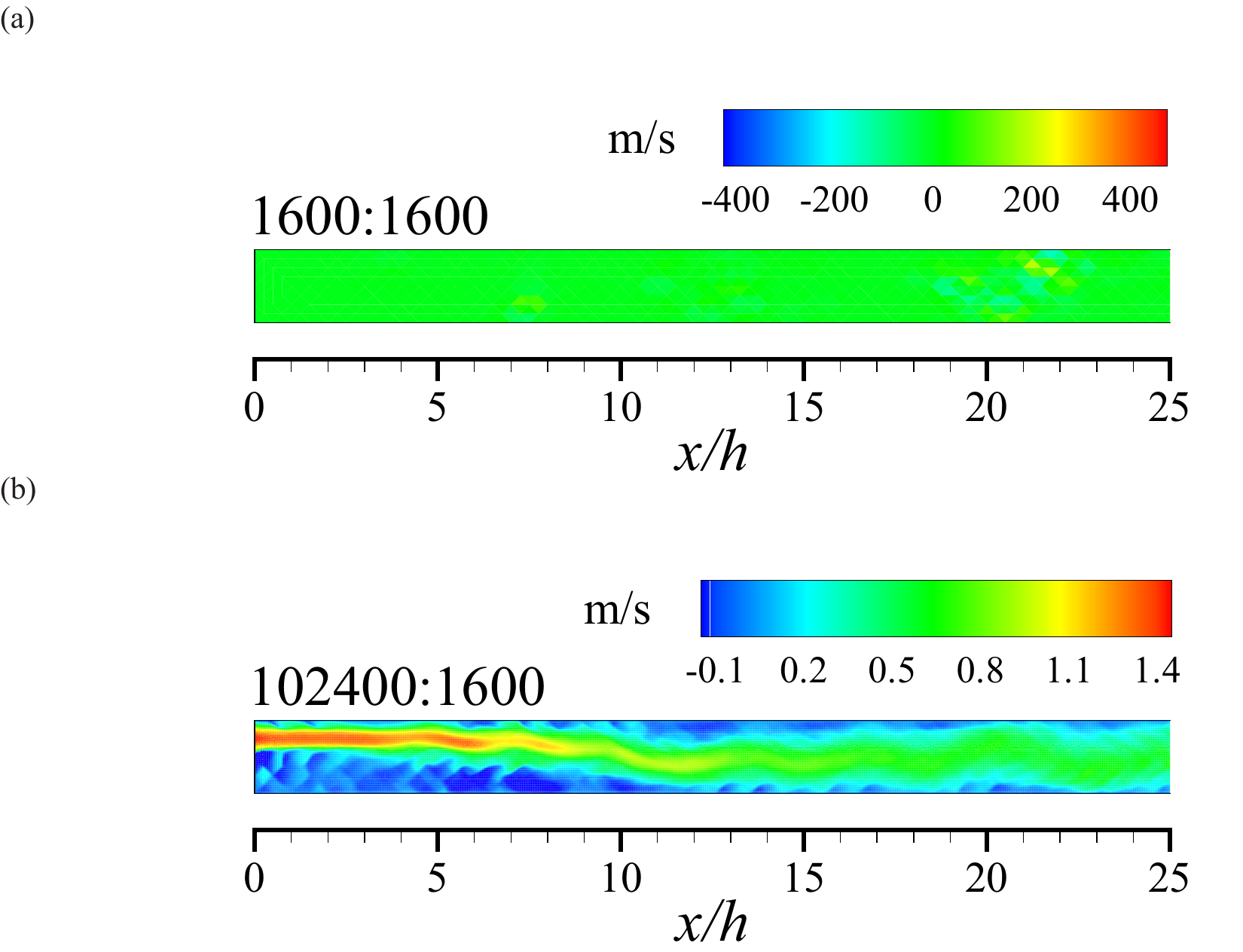}
\caption{Demonstration of partial mesh refinement application of the CGP method for the backward-facing step flow at $Re=800$, a comparison between axial velocity contours obtained using \textbf{a} Coarse scale computations (1600:1600), diverged, and \textbf{b} The CGP mesh refinement tool (102400:1600), converged. This figure is reproduced from Ref. [7].}
\label{fig:1}
\end{figure*}

Let's consider a condition that the standard numerical simulation diverged for the $N:N$ case due to a relative high Reynolds number or too coarse a mesh. Or the results obtained with a $N:N$ grid resolution are not sufficiently resolved and a fluid field with more detailed information is needed. The standard approach to resolving these common issues in pressure-correction methods [1] is to refine both the advection-diffusion and the Poisson grids. In contrast with this approach, the CGP strategy suggests refining the advection-diffusion grid, without changing the resolution of the Poisson mesh. To be more precise from a terminology point of view, CGP does not propose a new mesh refinement method; however, it guides users to implement available mesh refinement techniques for the grids associated with the nonlinear equations.

From a mesh refinement application point of view, the cost increment factor of the computational CGP tool $(h_{cgp})$ is approximated by
\begin{equation}
h_{cgp} = \frac{C_{cgp}}{C_c}.
\end{equation}
Similarly, this factor for a regular triangulation refinement $(h_f)$ is conjectured to be
\begin{equation}
h_{f} = \frac{C_{f}}{C_c}.
\end{equation}
Based on the above discussion, $h_f$ is greater than $h_{cgp}$. This is mainly due to the factor of $4^l$ that multiples $C_p$ in Eq. (2). These results imply that mesh refinement using the CGP idea is more cost effective than the standard technique. Note that we analyzed the computational cost for finite-element discretizations. A similar discussion is valid for finite volume/difference discretizations [2, 3].

\section{Results and discussion}
Here, we describe the concept by showing two simple examples. Both examples are taken from one of our recent published studies [7], but another interpretation of the numerical results of these examples is discussed here.

\begin{table*}
\centering
\caption{Comparison of relative norm errors and $h_f/h_{cgp}$ between the standard and CGP mesh refinement tools for the backward-facing step flow at $Re=800$. The norm errors are taken from Ref. [7]. * indicates that the simulation diverges after 96 time steps.}
\label{tab:1}   
\begin{tabular}{llllll}
\hline\noalign{\smallskip}
Resolution & $\| \textbf{u}\|_{L^\infty(V)}$& \%decrease in error & $\|\textbf{u}\|_{L^2(V)}$  & \%decrease in error & $h_f/h_{cgp}$ \\
\noalign{\smallskip}\hline\noalign{\smallskip}
1600:1600 & * & - & * &- & - \\
102400:1600 & 4.48543E-5 & - & 2.23362E-5 & - & 29.534 \\
\noalign{\smallskip}\hline\noalign{\smallskip}
6400:6400 & 1.64502E-4 & - & 6.12495E-5 & - & - \\
102400:6400 & 1.43997E-5 & 1042.398 & 4.87393E-6 & 1156.675 & 28.613 \\
\noalign{\smallskip}\hline\noalign{\smallskip}
25600:25600 & 5.17352E-5 & - & 4.21246E-6 & - & - \\
102400:25600 & 6.69447E-6 & 672.805 & 7.88459E-7 & 434.265 & 19.924 \\
\noalign{\smallskip}\hline
\end{tabular}
\end{table*}

\begin{figure}
\centering
\includegraphics[width=84 mm]{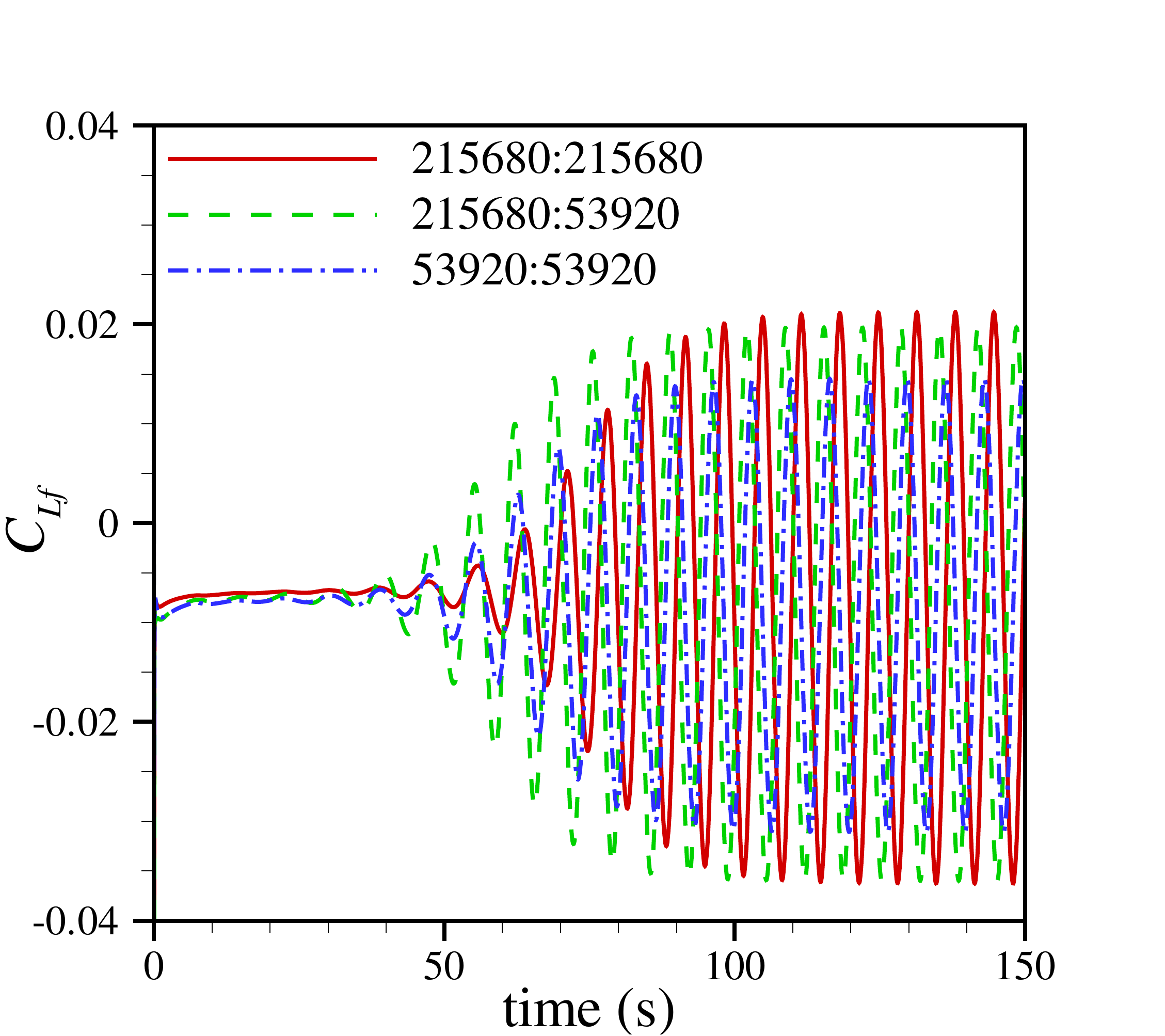}
\caption{Viscous lift coefficient obtained using the regular mesh refinement tool (215680:215680), the CGP mesh refinement tool (215680:53920), and the full coarse scale simulation (53920:53920). This figure is reproduced from Ref. [7].}
\label{fig:1}
\end{figure}

As the first example, consider the simulation of the flow over a backward facing step. Let's assume that one is interested in the flow information at the Reynolds number of  $Re=800$ (see Eq. (38) of Ref. [7] for the definition of the Reynolds number); however, due to wall clock time or computational resource limitations, he is not able to run a simulation with the required pure fine 102400:102400 grid resolution. On the other hand, because a coarse 1600:1600 resolution is not high enough, the simulation diverges after 96 time steps, as depicted in Fig. 1a. The CGP framework with an intermediate resolution of 102400:1600 provides a converged solution as shown in Fig. 1b. The relative velocity error norms with reference to the full fine simulation are of order 10E-5. Furthermore, the normalized reattachment length can be estimated around 14.0. Although the error percentage of this estimation is 16.67\% relative to that obtained by the standard computations, it is captured 30 times faster. Note that these results are achieved by only refining the advection-diffusion equation solver mesh. Table 1 lists relative norm errors of the velocity domain and $h_f/h_{cgp}$. For instance, refining the coarse mesh with the 6400:6400 spatial resolution using the CGP tool leads to a 1042.388\% reduction in the $L^2$ norm error of the velocity field, while it is 28.613 times cheaper than the regular mesh refinement technique. Note that in the case of 1600:1600 spatial resolution, because the simulation on the coarse grid diverges, there is no real number for $C_c$; however, if a virtual $C_c$ considered, $h_f/h_{cgp}=29.534$. 

As the second example, let's consider the flow over a circular cylinder computations described in Sect. 3.3 of Ref. [7]. The time evolution of the viscous lift coefficient $(C_{Lf})$ at the Reynolds number of $Re=100$ (see Eq. (39) of Ref. [7] for the definition of the Reynolds number) for three different combinations of the advection-diffusion and the Poisson grid resolutions is depicted in Fig. 2. Let's assume an exact measurement of the lift coefficient is needed for a specific engineering purpose. Using standard methods, this can be accomplished using either 215680:215680 or 53920:53920 grid resolutions. An implementation with the finer grid produces a more precise answer. It could be a user’s incentive to locally/globally refine the full coarse mesh. Obviously, this mesh refinement ends in an increase in CPU time for the simulation. In this case, our numerical experiments show that the increase is equal to 339271.6 s (over 94 hr). As discussed in Sect. 3.3 of Ref. [7], having a coarse mesh only degrades the level of accuracy of the viscous lift not the pressure lift. In fact, instead of refining the grids associated with both the nonlinear and linear equations, a mesh refinement of the nonlinear part is enough alone. Hence, in order to increase the precision of the lift force, one can refine the advection-diffusion grid and keep the resolution of the Poisson mesh unchanged. In this case, the CGP grid refinement cost factor is $h_{cgp}=3.638$, whereas this factor for the regular mesh refinement is $h_f=11.088$, illustrating a considerable saving of computational resources. 

As a last point, obviously the types of two dimensional flow simulations described here are not challenging computation problems today. These problems are merely used as examples to explain one of the features of the CGP algorithm. Practical applications of the CGP mechanism as a mesh refinement tool are expected to be useful for three dimensional flow simulations on parallel machines.

%and subsection a unique label (see Sect.~\bibitem{Ref31}).
%\paragraph{Paragraph headings} Use paragraph headings as needed.



\end{document}